\documentclass[sigconf,nonacm]{acmart}
\setcopyright{none}
\usepackage{booktabs}
\usepackage{graphicx}

\title{Train Often, Deploy Selectively: Forward-Gated Model Replacement in
Crypto Markets}

\author{Aditya Dutta}
\correspondingauthor
\orcid{0009-0004-7268-3629}
\email{aditya.dutta@emory.edu}
\affiliation{%
  \institution{Emory University}
  \city{Atlanta}
  \state{Georgia}
  \country{USA}
}

\renewcommand{\shortauthors}{Dutta}


\newcommand{\PooledCalendar}{0.1472\%}
\newcommand{\PooledCalendarCI}{0.1139\%--0.1754\%}

\newcommand{\PooledBlind}{0.0755\%}
\newcommand{\PooledBlindCI}{0.0521\%--0.0980\%}

\newcommand{\PooledMaintenance}{0.0428\%}
\newcommand{\PooledMaintenanceCI}{0.0301\%--0.0554\%}

\newcommand{\CanonicalPromotions}{114}

\newcommand{\CanonicalDecisions}{528}
\newcommand{\BreadthCalendar}{0.0821\%}
\newcommand{\BreadthBlind}{0.0458\%}
\newcommand{\BreadthMaintenance}{0.0270\%}
\newcommand{\TrialOneCalendar}{0.1738\%}
\newcommand{\TrialOneBlind}{0.1036\%}
\newcommand{\TrialOneMaintenance}{0.0470\%}

\newcommand{\TrialOneSpearman}{0.524}
\newcommand{\TrialOneAgreement}{74.0\%}

\newcommand{\TrialTwoCalendar}{0.1760\%}
\newcommand{\TrialTwoBlind}{0.0751\%}
\newcommand{\TrialTwoMaintenance}{0.0491\%}

\newcommand{\TrialThreeCalendar}{0.1683\%}
\newcommand{\TrialThreeBlind}{0.0899\%}
\newcommand{\TrialThreeMaintenance}{0.0414\%}

\begin{document}
\begin{abstract}
Production forecasting systems retrain models regularly, but a retrained candidate does not necessarily outperform a continuously maintained incumbent that has continued to learn. We introduce \emph{Shadow Before Swap} (SBS), a deployment policy that warm-refits a challenger off the serving path, evaluates it against the maintained incumbent on the same next week of delayed labels, and promotes it only after a fixed paired negative-log-likelihood (NLL) advantage.

In historical replay over two nonoverlapping Binance episodes spanning 48 UTC weeks, three seeds, eight underlyings, and two perpetual-futures contract types, SBS reduces NLL by \PooledCalendar{} relative to calendar replacement, \PooledBlind{} relative to schedule-matched automatic promotion, and \PooledMaintenance{} relative to continuous maintenance. The corresponding episode-stratified four-week block intervals are \PooledCalendarCI{}, \PooledBlindCI{}, and \PooledMaintenanceCI{}. SBS promotes \CanonicalPromotions{} of \CanonicalDecisions{} challengers, reducing deployed model changes by 78.4\% while improving the serving trajectory. The effect remains directionally consistent across seeds, trial budgets, promotion margins, an earlier 20-asset panel, and a topology-matched supervised objective. SBS thus provides a practical deployment policy that improves probabilistic forecasts while limiting consequential model-state transitions.
\end{abstract}

\maketitle

\section{Introduction}
Production forecasters are maintained systems, not static checkpoints. Between scheduled rebuilds, a serving model may update normalization statistics, consume labels as they mature, and adapt its prediction head. A recent-window refit can therefore beat an old checkpoint while losing to the state that is actually serving. Deploying that refit also changes the initialization of every later update and challenger. The operational question is consequently not just \emph{when to retrain}, but \emph{whether this trained state should replace this maintained state}.

That distinction is easy to miss in an offline benchmark. A newly fitted model is usually compared with the checkpoint from which the experiment began. In a live system, however, the relevant alternative is the incumbent at the moment of release, after it has absorbed another week of inputs and mature labels. The release decision also has lasting consequences: the selected state becomes the parent of the next refit. Model replacement is therefore a sequential control problem, even when the underlying learner and release cadence are fixed.

Production-ML literature emphasizes validation, pipeline integrity, monitoring, versioned artifacts, and operational risk controls \cite{breckmltest,tfx,sculleydebt,nistairmf}. Current banking model-risk guidance likewise separates validation, implementation, ongoing monitoring, and governance \cite{fedmodelrisk2026}. Canary analysis evaluates a bounded live production change before roll-forward or rollback \cite{beyercanary}. These controls are difficult for financial forecasts: labels are delayed, observations are serially dependent, and a useful incumbent may continue adapting during evaluation. A deployment gate should compare candidate and incumbent under the same market shocks without freezing the incumbent or exposing the candidate to serving.

SBS implements that comparison as a causal shadow trial. At each scheduled boundary it clones the complete incumbent, warm-fits the clone using mature recent history, and advances challenger and incumbent independently on the same next week. Once the week's labels mature, SBS promotes only if mean challenger NLL is at least $10^{-4}$ lower. The continuously maintained state is the deployment null. A schedule-matched blind policy trains and shadows the same challengers but always promotes, isolating the value of trial evidence from the value of waiting.

Across 48 nonoverlapping weeks, SBS beats immediate calendar replacement and blind promotion. It also improves by \PooledMaintenance{} over a continuously maintained incumbent. This last comparison rules out a trivial explanation in which the safest policy is simply never to refit. SBS accepts 21.6\% of proposals, captures the aggregate value of those refits, and avoids 414 unnecessary model changes.

This paper makes three contributions:
\begin{itemize}
\item It defines model replacement as authorization against a maintained, recursively evolving incumbent, rather than evaluation against a stale
checkpoint.
\item It gives a delayed-label paired gate and a schedule-matched blind contrast that identify the value of forward evidence.
\item It provides replicated time- and asset-breadth evidence that selective deployment improves proper probability scores while materially reducing replacement frequency.
\end{itemize}
\paragraph{Why this matters.}
Existing research largely asks how to train better forecasting models. Production systems face an additional decision: whether a newly trained state should replace a continuously adapting incumbent. The problems differ because deployment changes the parent of every future update and challenger. SBS formalizes this release-authorization problem independently of forecasting architecture. \emph{The contribution is a deployment policy rather than a forecasting architecture.} It can sit between an existing retraining pipeline and model registry: training continues on schedule, but installation requires evidence collected after fitting.

\begin{figure}[t]
\centering
\scriptsize
\begin{tabular}{c@{\,$\rightarrow$\,}c@{\,$\rightarrow$\,}c@{\,$\rightarrow$\,}c@{\,$\rightarrow$\,}c}
\fbox{\strut Train clone} &
\fbox{\strut Shadow} &
\fbox{\strut Labels mature} &
\fbox{\strut Compare NLL} &
\fbox{\strut\shortstack{Promote\\or retain}}
\end{tabular}
\vspace{3pt}

\begin{minipage}{0.98\columnwidth}
\small
\textbf{Implementation procedure}\\[-3pt]
\begin{enumerate}
\setlength{\itemsep}{0pt}
\setlength{\parsep}{0pt}
\item Deep-copy the incumbent model, normalizer, delayed-label queue, and online-update state.
\item Fit $(\mu',\sigma')$ on mature history and set $W'=W\,\mathrm{diag}(\sigma'/\sigma)$, $b'=b+W((\mu'-\mu)/\sigma)$; warm-fit with fresh, training-only Adam.
\item Advance isolated incumbent and challenger branches through the paired trial, including their separate queues and mature-label updates.
\item After labels mature, promote if Eq.~\ref{eq:value} clears $\tau$; otherwise retain the incumbent. Continue the selected state recursively.
\end{enumerate}
\end{minipage}
\caption{\textbf{SBS decision flow and pseudocode.} The challenger remains off the serving path until forward labels mature. Deep copies prevent shared mutable state; the deployed head updater is momentum-free SGD.}
\Description{A five-stage flow diagram and four implementation steps describe training, shadow evaluation, label maturation, comparison, and recursive promotion or retention.}
\label{fig:pseudocode}
\end{figure}

\section{Forward-Gated Replacement}\label{sec:method}
\subsection{Maintained state and paired trial}
Let the deployable state be $S_t=(\phi_t,W_t,N_t,Q_t,O_t)$: representation, prediction head, causal normalizer, delayed-label queue, and online optimizer. A maintenance operator $\mathcal M_t$ releases labels available by $t$, applies their head updates, predicts the current example, enqueues its delayed label, and then refreshes input statistics:
\begin{equation}\label{eq:maintenance}
S_{t+1}=\mathcal M_t(S_t;x_t,\{(x_i,y_i):a_i\leq t\}),
\end{equation}
where $a_i$ is label-availability time. Prediction precedes use of the current label, following prequential evaluation \cite{dawidprequential} and delayed online learning \cite{joulani,labeldelay}.

At action block $b$, challenger $C_b$ is cloned from incumbent $I_b$ and fitted only on history mature at the training boundary. Both states then advance on the same trial stream $\mathcal T_b$. At decision time $d_b$,
\begin{equation}\label{eq:value}
\widehat v_b=\frac{1}{n_b}\sum_{i\in\mathcal T_b:a_i\leq d_b}
\left[\ell(p_{I_b,i},y_i)-\ell(p_{C_b,i},y_i)\right].
\end{equation}
SBS promotes when
\begin{equation}\label{eq:gate}
\widehat A_b=\mathbf 1\{\widehat v_b\geq\tau\},\qquad \tau=10^{-4}.
\end{equation}
Positive $\widehat v_b$ means lower challenger NLL. Scoring both states on the same observations yields paired loss differentials, as in comparative predictive-ability analysis \cite{dieboldmariano,giacominiwhite}, so common market shocks enter both losses; temporal dependence nevertheless remains.

A \emph{challenger} is one contract-underlying-seed state trained at one scheduled boundary. Equation~\ref{eq:gate} produces one promotion decision for that state. The threshold is applied to its paired trial mean \emph{before} cross-asset or cross-contract aggregation. Aggregation is used only to evaluate complete policies.

\subsection{Causal timing and state isolation}
Each decision follows the same sequence. At the scheduled boundary, the incumbent first processes every label whose availability time has passed. The challenger is then cloned and fitted using only examples mature at that boundary. Predictions from both branches are recorded during the following trial, but neither branch can use a trial label before its 300-second delay expires. The gate is evaluated only after all eligible trial labels have matured. This ordering prevents training-window overlap from leaking future labels into the release decision.

The two branches see the same examples in the same order, yet their mutable states remain separate. Each branch carries its own normalizer and delayed queue; head updates on one branch cannot alter the other. A rejected challenger is discarded in full. An accepted challenger contributes its fitted parameters, normalizer, and accumulated trial updates, and then becomes the parent of later challengers. The replay follows the same ownership rule as a production champion--challenger service, which is essential for measuring the effect of a release policy rather than a collection of isolated model comparisons.

\subsection{Comparators and recursive estimand}
Four complete replays identify the source of value. \emph{Maintenance} applies Eq.~\ref{eq:maintenance} but never full-refits. \emph{Calendar} warm-refits and deploys immediately at each boundary. \emph{Blind} follows the SBS training and trial schedule but always promotes. \emph{SBS} promotes only under Eq.~\ref{eq:gate}. SBS versus blind isolates authorization; SBS versus maintenance tests whether accepted refits add value; SBS versus calendar tests the operational default of scheduled replacement.

Calendar and blind promotion are intentionally distinct. Calendar begins serving the refit as soon as training ends, whereas blind promotion waits through the same shadow interval as SBS and then installs every challenger. Their difference measures the cost or benefit of waiting one week even when forward evidence is ignored. The SBS--blind contrast then changes only the authorization decision. Maintenance supplies a harder null: it keeps the incumbent's online adaptation but never refreshes the full representation. Together, the three contrasts separate timing, authorization, and the value of accepted refits.

After policies take different actions, their descendants diverge:
\begin{equation}\label{eq:recursive}
S_{b+1}^{\pi}=G(S_b^\pi,A_b^\pi,\omega_b).
\end{equation}
The estimand is therefore the loss contrast between full recursive serving trajectories, not classification accuracy over a fixed challenger set. This is also why each margin sensitivity is replayed from start to finish rather than applied retrospectively to one set of decisions.

\section{Experimental Design}\label{sec:design}
\subsection{Evidence chronology}
Rule development used 26 UTC weeks from February 3 through August 3, 2025, with seed 0. The one-week trial, $10^{-4}$ deadband, policy family, aggregation order, mature-label scoring, and temporal-block analysis were then frozen in the versioned repository before the next period was analyzed. Repository configuration and run manifests make the lock reproducible. We use the precise term \emph{development-selected and frozen}: the evidence supports temporal separation between rule selection and subsequent evaluation, without claiming independent preregistration.

The first out-of-time episode spans August 4, 2025 through January 4, 2026 (22 weeks). A second, nonoverlapping episode spans January 5 through July 5, 2026 (26 weeks). Both use three seeds, eight underlyings, and Binance USD-M and COIN-M perpetual contracts. The second period had previously supported a maintenance audit, but its recursive SBS outcomes were unopened when the robustness protocol was registered. We therefore classify it precisely as a \emph{retrospective historical robustness episode}; its unopened recursive policy outcomes still provide a nonoverlapping test of the frozen rule. An earlier March--June 2024 panel tests asset breadth with 20 USD-M underlyings. Objective, cadence, Temporal-CNN, and Coinbase analyses are secondary stress tests.

The $10^{-4}$ threshold is an operational deadband selected during development, equivalent to about 0.009\% of the baseline weekly NLL. It requires a positive advantage larger than numerical ties while remaining well below the observed penalties of rejected challengers. Section~\ref{sec:results} reports complete recursive replays at zero, $10^{-4}$, $2.5\times10^{-4}$, and $5\times10^{-4}$; the conclusion is stable throughout that range. The primary threshold should thus be read as a conservative release tolerance, not an estimate of a universal market constant. An organization with a larger rollback or validation cost could set a wider deadband before evaluation.

\subsection{Markets, data, target, and missingness}
The primary archive is Binance public futures data \cite{binancedata}. The panel contains BTC, ETH, BNB, XRP, SOL, DOGE, ADA, and BCH in USD-M and COIN-M forms. These are two contract types on one exchange, not independent venues. USD-M and COIN-M preprocessing has identical economic features after explicit unit conversion: log base quantity, log USD notional, and within-side depth shares. Contract sizes are taken from hashed exchange metadata.

Complete ten-band depth observations are snapped to a 30-second UTC grid and deduplicated. A day must contain at least 95\% of its 2,880 scheduled observations. One isolated missing grid step may be causally forward-filled; longer gaps remain discontinuities, and no 60-snapshot input or 300-second target window may cross them. Missing registered days are omitted consistently from every policy state.

Each input has 60 snapshots and 40 depth-band features. The target is down, neutral, or up according to the 300-second future transaction-price return outside a fixed five-basis-point neutral band. Reference prices use only the latest completed official one-minute futures close available at prediction time. The 22-week episode contains 6,319,220 scored stream examples with class shares 34.75\%/30.88\%/34.37\% (down/neutral/up); the 26-week episode contains 6,789,702 with shares 33.29\%/33.97\%/32.74\%. The earlier 20-asset panel contains 5,536,506 with shares 35.97\%/27.98\%/36.05\%.

The Coinbase screen uses different preprocessing. It matches the 60-by-40 shape, 30-second decision cadence, five-basis-point band, and 300-second horizon, but causally downsamples exchange snapshots and labels midpoint returns rather than Binance minute-close returns. The screen contains 55,367 trial examples. Its down, neutral, and up shares are 41.17\%, 16.84\%, and 41.98\%; it is used as a focused one-action transport screen. This targeted screen asks whether the decision interface survives a change of venue and price construction; it complements the recursive Binance evidence.

\subsection{Models and maintenance}
The primary network projects each snapshot to 24 channels, pools temporal mean, final, and maximum features, applies a second 24-unit layer, and uses a linear three-class head. Layer-local Forward--Forward training fits the representation \cite{hintonff}; mature labels supervise the head. A topology- and capacity-matched cross-entropy network tests transfer across training objectives. A 32-channel dilated residual Temporal-CNN supplies a deliberately difficult failure-containment stress test.

All policies refresh causal normalizer moments from current unlabeled inputs at rate 0.01. Labels mature after 300 seconds and update only the head by plain SGD at $3\times10^{-4}$. Warm refits use the preceding 28 calendar days: 21 for training and seven for validation, with at least 14 source days. They fit a fresh normalizer, apply the function-preserving reparameterization in Figure~\ref{fig:pseudocode}, and train with fresh Adam at $10^{-3}$, batch size 128, at most 50 epochs, and validation-NLL patience three.

NLL accumulation floors true-class probability at $10^{-300}$ only to prevent floating-point log underflow. Paired gate NLL is computed directly from logits with stable log-sum-exp. No reported primary prediction approaches the floor; the floor therefore does not truncate the observed policy contrasts.

\subsection{Replay integrity}
Every policy starts an episode from the same checkpoint and consumes the same ordered market stream. UTC boundaries determine training, validation, and label availability independently of policy actions. Run manifests bind prepared tensors, exchange metadata, configuration, and starting checkpoints by content hash. Challenger--incumbent trials and complete policy replays are therefore paired on the same exogenous tape even after their states diverge.

\subsection{Aggregation and uncertainty}
Daily sufficient statistics are pooled to stream-week NLL. Seeds are averaged first within stream-week, underlyings equally within contract type, the two contract types equally within UTC week, and weeks equally over time. This is an equal decision-budget estimand: high-volume assets cannot dominate by supplying more predictions.

For policies $\pi$ and $q$, the reported relative effect is
\begin{equation}\label{eq:effect}
100\left(1-\frac{L_{\pi}}{L_q}\right),
\end{equation}
where $L$ is the hierarchically aggregated NLL. Positive values favor SBS. The aggregation order and weights are fixed before resampling, so a bootstrap draw changes the sampled sequence of weeks without changing the intended cross-sectional estimand.

The primary uncertainty summary uses 10,000 deterministic circular moving-block resamples with four-week blocks \cite{politisromano,lahiriresampling}. Two-, six-, and eight-week blocks, stationary bootstrap, Newey--West lag-four intervals, contract exclusions, and leave-one-asset-out estimates are sensitivities \cite{stationarybootstrap,neweywest}. The 48-week analysis resamples within each episode and then combines effects with fixed 22/48 and 26/48 weights. Pointwise intervals are reported for each named contrast; a simultaneous max-deviation interval addresses the family of three canonical comparators.

\section{Empirical Results}\label{sec:results}
\subsection{Selective deployment improves the serving trajectory} Table~\ref{tab:canonical} reports the main result. SBS improves relative NLL against calendar, blind promotion, and maintenance in each episode and in the episode-stratified combination. The pooled estimates are \PooledCalendar{}, \PooledBlind{}, and \PooledMaintenance{}; in absolute NLL they are 0.001567, 0.000803, and 0.000455 per forecast, respectively. The four-week block intervals describe temporal variation conditional on each realized recursive replay. Under that interpretation, every compatible range in Table~\ref{tab:canonical} is positive.

\begin{table}[t]
\caption{\textbf{Canonical SBS effects.} Relative NLL reductions (\%) with pointwise 95\% four-week circular moving-block intervals. The 48-week result resamples within episodes and uses fixed 22/26-week weights.}
\label{tab:canonical}
\centering
\small
\setlength{\tabcolsep}{2.5pt}
\begin{tabular}{lccc}
\toprule
Evidence & Calendar & Blind & Maintenance \\
\midrule
22 weeks & 0.1157 & 0.0423 & 0.0379 \\
 & [0.0676, 0.1573] & [0.0146, 0.0689] & [0.0242, 0.0533] \\
26 weeks & 0.1738 & 0.1036 & 0.0470 \\
 & [0.1265, 0.2103] & [0.0673, 0.1380] & [0.0263, 0.0671] \\
48 weeks & 0.1472 & 0.0755 & 0.0428 \\
 & [0.1139, 0.1754] & [0.0521, 0.0980] & [0.0301, 0.0554] \\
\bottomrule
\end{tabular}
\end{table}

\begin{figure*}[t]
\centering
\includegraphics[width=0.99\textwidth]{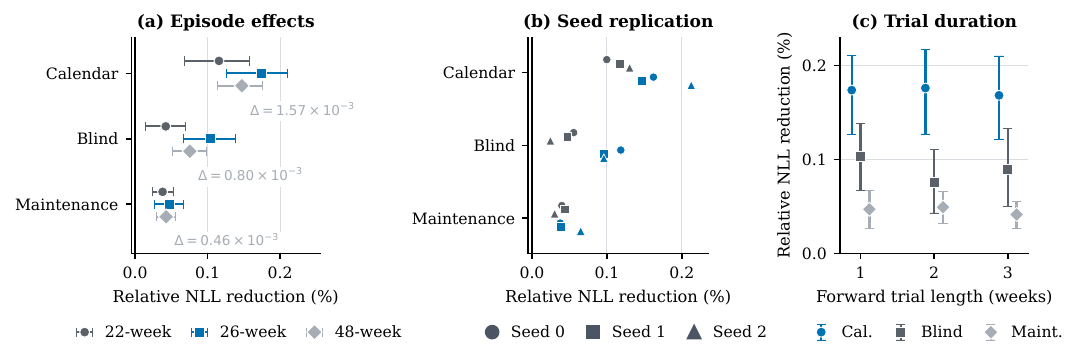}
\caption{\textbf{Selective replacement replicates.}
(a) Episode and stratified relative effects; green annotations give exact pooled absolute NLL reductions in $10^{-3}$ units. Whiskers are pointwise 95\% four-week block intervals. (b) All seed-specific effects are positive in both episodes. (c) Discrete one-, two-, and three-week trial budgets remain positive; points are intentionally not joined because duration is a discrete experimental choice.}
\Description{Panel a shows positive forest-plot effects and absolute effect labels. Panel b shows positive seed effects. Panel c shows unjoined trial duration effects with confidence intervals.}
\label{fig:replication}
\end{figure*}

\paragraph{Reading the three contrasts.}
The ordering of the effects is informative. The largest gap, against calendar replacement, reflects both the value of waiting for forward labels and the value of acting on them. Blind promotion pays the same one-week delay, making its \PooledBlind{} gap the cleanest measure of authorization. The \PooledMaintenance{} improvement is smaller because maintenance is already a strong adaptive baseline, but its positive sign shows that SBS does more than suppress change: accepted refits improve on the online-updated incumbent.

\paragraph{Why NLL improvements matter.}
Effects are measured per prediction, whereas NLL is additive over forecasts. The 0.000455 absolute maintenance gain therefore accumulates to about 455 total log-loss units per million predictions. At matched 5\% coverage in the 22-week episode, the calendar contrast corresponds to roughly 200 additional correct high-confidence decisions per million opportunities. These are service-scale gains in the full probability vector, not rounded changes in top-class accuracy.

Operational scale reinforces the forecast result. SBS makes 114 promotions among 528 proposals, avoiding 414 artifact validations, staged releases, and potential rollback events relative to blind promotion. Challenger training and a bounded shadow pass remain the evaluation cost; deployed-state turnover falls by 78.4\%. The measured joint outcome is therefore stronger probability forecasts with substantially fewer consequential state transitions.

The gain is not confined to one seed or a single favorable week. All 18 episode-seed-comparator effects are positive. Weekly median policy differences are positive for every comparator in both episodes. The worst weekly relative differences (22-week/26-week) are $-0.0232\%/0.0000\%$ versus calendar, $-0.0354\%/-0.0593\%$ versus blind, and $-0.0080\%/-0.0184\%$ versus maintenance; 17/22 and 22/26 weeks favor SBS over calendar, 13/22 and 21/26 over blind, and 17/22 and 21/26 over maintenance. Thus gains vary over time but are not generated by one isolated week. Brief weekly reversals are the expected local cost of operating in a nonstationary stream; positive medians, majorities of favorable weeks, and block intervals above zero show that they do not drive the aggregate conclusion.

\subsection{Threshold and evidence-budget sensitivity}
Table~\ref{tab:margins} reports every margin replay with its full pointwise interval. All four margins improve on all three comparators in both episodes. The near-flat policy effects show that the conclusion does not depend on the development-selected $10^{-4}$ value. Increasing the deadband primarily reduces replacement frequency: promotions fall from 56 to 42 in the 22-week episode and from 66 to 55 in the 26-week episode.

\begin{table*}[t]
\caption{\textbf{Full promotion-margin sensitivity.} Relative NLL reductions (\%) with pointwise 95\% four-week block intervals. P/R means the number of promotions/rejections among mature challenger decisions.}
\label{tab:margins}
\centering
\small
\setlength{\tabcolsep}{3pt}
\begin{tabular}{ccclll}
\toprule
Episode & $\tau$ & P/R & Calendar & Blind & Maintenance \\
\midrule
22 weeks & 0 & 56/184 & 0.1174 [0.0693, 0.1591] &
0.0441 [0.0165, 0.0704] & 0.0397 [0.0258, 0.0548] \\
& $10^{-4}$ & 50/190 & 0.1157 [0.0676, 0.1573] &
0.0423 [0.0146, 0.0689] & 0.0379 [0.0242, 0.0533] \\
& $2.5{\times}10^{-4}$ & 47/193 & 0.1155 [0.0662, 0.1589] &
0.0421 [0.0110, 0.0712] & 0.0377 [0.0224, 0.0553] \\
& $5{\times}10^{-4}$ & 42/198 & 0.1135 [0.0650, 0.1562] &
0.0401 [0.0093, 0.0685] & 0.0357 [0.0213, 0.0520] \\
\midrule
26 weeks & 0 & 66/222 & 0.1742 [0.1266, 0.2113] &
0.1039 [0.0671, 0.1388] & 0.0473 [0.0266, 0.0680] \\
& $10^{-4}$ & 64/224 & 0.1738 [0.1265, 0.2103] &
0.1036 [0.0673, 0.1380] & 0.0470 [0.0263, 0.0671] \\
& $2.5{\times}10^{-4}$ & 59/229 & 0.1701 [0.1235, 0.2064] &
0.0998 [0.0641, 0.1342] & 0.0432 [0.0233, 0.0624] \\
& $5{\times}10^{-4}$ & 55/233 & 0.1705 [0.1241, 0.2069] &
0.1002 [0.0645, 0.1351] & 0.0436 [0.0242, 0.0623] \\
\bottomrule
\end{tabular}
\end{table*}

Changing the evidence budget reaches the same conclusion (Figure~\ref{fig:replication}c). In the 26-week episode, calendar gains are \TrialOneCalendar{}, \TrialTwoCalendar{}, and \TrialThreeCalendar{} for one-, two-, and three-week trials; blind gains are \TrialOneBlind{}, \TrialTwoBlind{}, and \TrialThreeBlind{}; maintenance gains are \TrialOneMaintenance{}, \TrialTwoMaintenance{}, and \TrialThreeMaintenance{}. All corresponding pointwise four-week intervals remain positive. Two weeks has the lowest observed NLL, but that ranking was observed after outcomes and is not promoted to a new primary rule.

The margin and duration analyses answer different deployment questions. Changing $\tau$ alters how much measured advantage is required after a fixed trial; changing the trial length alters both the evidence available and the time before a decision can be acted upon. The stable sign across both axes is more informative than the identity of the best cell. It shows that the result survives reasonable changes to the release budget without turning the robustness grid into a second model-selection exercise.

\subsection{Forward evidence is useful without perfect ranking}
\begin{figure*}[t]
\centering
\includegraphics[width=0.99\textwidth]{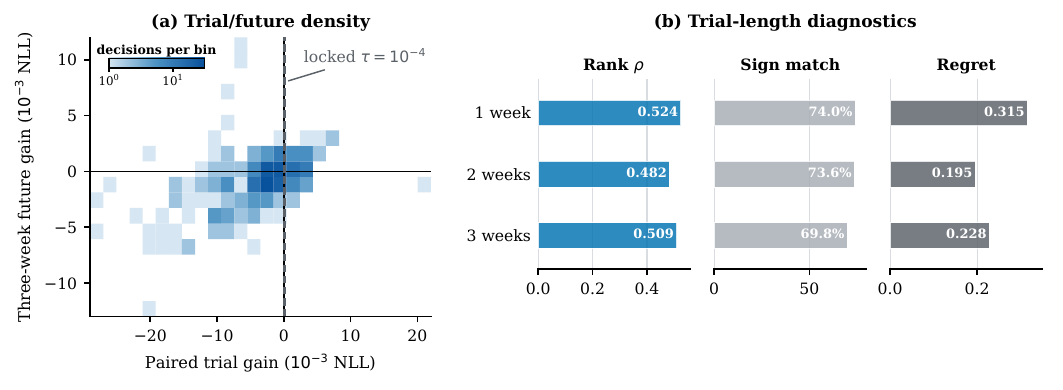}
\caption{\textbf{What the forward trial measures.}
(a) Binned density of one-week paired trial gain and three-week future challenger value across 288 decisions; color gives the number of decisions in each cell, and the dashed line is the development-selected, locked threshold. (b) Duration-specific rank association, sign agreement, and mean local hindsight regret on separate scales. Higher rank and sign agreement and lower regret are better. These are descriptive mechanism diagnostics, not policy-level confidence intervals; policy-level evidence comes from the complete recursive replays.}
\Description{A binned density heatmap shows association between trial and future value. Aligned horizontal-bar plots report rank correlation, sign agreement, and regret at three discrete trial durations.}
\label{fig:trialfuture}
\end{figure*}

The canonical 26-week episode contains 288 mature challenger decisions. Paired trial gain has Spearman association \TrialOneSpearman{} with value over the next three mature weeks and \TrialOneAgreement{} sign agreement. Figure~\ref{fig:trialfuture}a replaces an overplotted scatter with binned density. Panel b reports rank association, sign agreement, and local hindsight regret as aligned small multiples because they have different units. Longer trials change which descendants are reached, while rank, sign, and regret remain in a similar operating range. The policy result therefore does not rely on choosing the best-looking duration after observing outcomes. 

Perfect ranking is unnecessary for a useful gate. A one-week trial is asked to screen out damaging state transitions, not to recover the exact ordering of three-week challenger value. Moderate rank association can be sufficient when the lower tail contains costly refits and useful challengers retain the opportunity to clear the threshold. The recursive policy results provide the relevant test: screening errors are already reflected in later descendants and in the final NLL trajectory.

\subsection{Time and asset breadth, objective transfer, and safety}
\begin{figure}[t]
\centering
\includegraphics[width=\columnwidth]{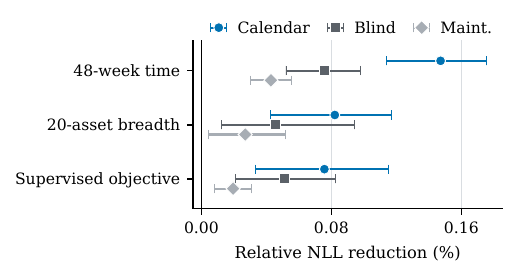}
\caption{\textbf{Breadth evidence.}
Primary time breadth, earlier 20-asset breadth, and topology-matched objective transfer with pointwise 95\% four-week block intervals.}
\Description{A forest plot shows positive effects across time, asset, and objective evidence.}
\label{fig:breadth}
\end{figure}

The earlier 14-week, 20-asset USD-M panel improves by \BreadthCalendar{} over calendar, \BreadthBlind{} over blind, and \BreadthMaintenance{} over maintenance. Four-week intervals are 0.0427\%--0.1170\%, 0.0121\%--0.0940\%, and 0.0044\%--0.0515\%, respectively. A topology-matched cross-entropy challenger also remains positive: 0.0757\%, 0.0508\%, and 0.0195\% against the same comparators. We describe these as time and asset breadth and training-objective transfer, not broad
``generalization.'' SBS also beats automatic calendar replacement at tested two-, four-, and eight-week cadences by 0.1195\%, 0.0997\%, and 0.0580\%. Figure~\ref{fig:breadth} collects the three breadth contrasts.

\begin{figure*}[t]
\centering
\includegraphics[width=0.99\textwidth]{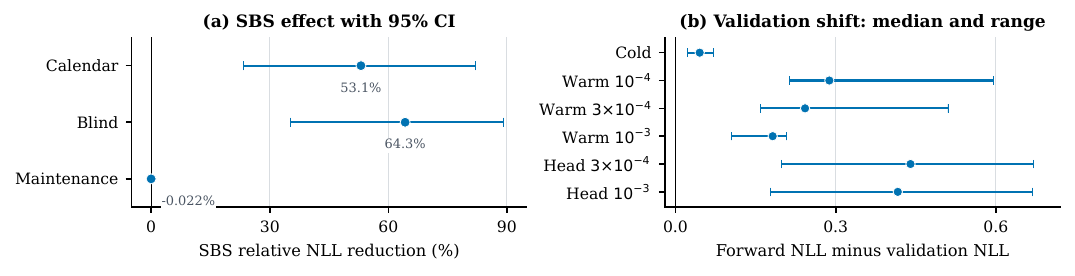}
\caption{\textbf{Temporal-CNN failure containment.}
(a) Bootstrap-median SBS relative NLL reduction with pointwise 95\% four-week block intervals in the common-recipe stress case. (b) Median and range of forward-minus-validation NLL for six diagnostic refit configurations (six market-seed cells each).}
\Description{Two equal-sized panels show SBS effects against three deployment baselines with confidence intervals and positive validation-to-forward CNN loss gaps.}
\label{fig:cnnsafety}
\end{figure*}

The Temporal-CNN experiment isolates the safety value of authorization under a severely misspecified candidate pipeline. Applying the common refit recipe to this architecture produces large validation-to-forward NLL gaps across warm, head-only, and cold diagnostic fits (Figure~\ref{fig:cnnsafety}b). Recursive automatic deployment then amplifies the mismatch: calendar and blind NLL reach 12.6506 and 15.8667, versus 1.07117 for maintenance. SBS rejects 236 of 240 proposals and reaches 1.07141. The gate therefore retains 99.98\% of maintenance performance while preventing the severe loss escalation caused by automatic replacement.

The $-0.0221\%$ maintenance contrast is the measured safety cost of evaluating a candidate generator that supplies almost no deployable refits, not evidence that the gate fails to discriminate: 236 of 240 proposals are rejected. Positive validation-to-forward gaps diagnose the mismatch before the policy comparison, and calendar and blind promotion demonstrate the consequence of installing those states. The stress test establishes a distinct result from architecture quality: forward authorization can keep a serving trajectory near its strong incumbent even when conventional validation repeatedly favors unsafe replacements.

NLL is a strictly proper probability score \cite{gneitingraftery}. In the 22-week secondary matched-coverage analysis, SBS also lowers Brier score \cite{brier1950} by 0.000678 versus calendar, 0.000370 versus blind, and 0.000274 versus maintenance. Following risk--coverage evaluation in selective classification \cite{chowreject,elyanivwiener,geifmanselective}, equal-coverage comparisons favor SBS over calendar and maintenance, showing that the gain is not generated by greater abstention. These proper-score and matched-coverage gains establish forecast value. Converting them into executable profit is a separate strategy-level estimand requiring prespecified trading costs, fills, and market impact.

\section{Inference and Interpretation}
\subsection{What the block intervals quantify}
The inferential sample is 48 aggregate UTC weeks; the millions of forecasts determine service-scale impact but are not treated as independent evidence. Forecast windows overlap, assets share market shocks, and an early action affects later descendants. We therefore average seeds first, fix cross-sectional weights, and resample adjacent blocks of aggregate weeks. Bootstrap draws recompute NLL ratios. Simultaneous 95\% lower endpoints for the three pooled contrasts are 0.1160\%, 0.0443\%, and 0.0116\%, so the family conclusion remains positive after allowing selection among the baselines.

These intervals target temporal robustness of the realized recursive trajectories: they ask whether the advantage is distributed across market weeks. Regenerating states after reordering weeks would instead retrain on synthetic chronologies, changing queues, normalizers, and every descendant; that is a different intervention, not a refinement of the observed-policy bootstrap. Counterfactual-history uncertainty requires independent episodes or a validated market simulator. Accordingly, block inference is paired with two nonoverlapping episodes, three seeds, and the earlier asset panel.

\subsection{Statistical and practical significance}
The block intervals measure stability across market time; promotion counts measure how often the deployed state changes to obtain that advantage. Here both favor SBS: NLL improves against maintenance and automatic replacement while releases fall sharply. Because NLL is strictly proper, the effect concerns the full probability vector. Its additive service-scale value and the matched-coverage diagnostic support practical significance; monetary value requires a specified downstream decision rule.

\subsection{Operational interpretation and scope}
SBS separates candidate-generation recall from deployment precision. Any training system may propose aggressively; the gate controls which state transition and descendants enter service. This matches production and financial model-risk guidance that distinguishes validation, release, monitoring, recovery, and change management \cite{breckmltest,tfx,nistairmf,fedmodelrisk2026,beyercanary}. SBS addresses predeployment authorization. It does not replace postdeployment drift monitoring or rollback; it supplies the missing decision between validation and those established release controls.

\begin{table}[t]
\caption{\textbf{Incremental operational footprint of SBS.}}
\label{tab:cost}
\centering
\small
\setlength{\tabcolsep}{4pt}
\begin{tabular}{ll}
\toprule
Component & Additional cost \\
\midrule
Challenger inference & One bounded shadow pass \\
Storage & One versioned challenger state \\
Decision latency & One forward-trial week \\
Online serving & Incumbent path unchanged \\
\bottomrule
\end{tabular}
\end{table}

Table~\ref{tab:cost} makes the deployment trade explicit. The registry records the challenger's parent, training window, preprocessing identity, trial, and cutoff; promotion then uses ordinary canary, monitoring, and rollback controls. Equal asset and contract weighting is the deliberate release-policy estimand. Capital- or risk-weighted deployment requires weights frozen before outcomes and complete replay, preventing post-hoc portfolio choices from changing the governance conclusion.

The shadow branch consumes the same feature stream as the incumbent but never enters the latency-critical serving path. Only one challenger must be retained, and rejection permits its state to be discarded after the audit record is written. Thus SBS does not increase online inference latency or change the existing retraining cadence. Its principal operational charge is the deliberate one-week decision delay. That charge is visible and tunable, unlike the downstream cost of an unnecessary deployment: artifact review, staged rollout, monitoring, and possible rollback recur for every promoted state. The observed 414 avoided promotions therefore represent repeated workflow savings in addition to lower model turnover; their monetary value remains organization specific.

\subsection{Limitations}
Evidence comes primarily from Binance perpetual futures and one incumbent maintenance mechanism; the 20-asset, Coinbase, supervised, and CNN panels test specific dimensions rather than universal transport. We do not claim that the development-selected threshold is universally optimal or that forecast-score gains are direct trading profits. The estimand is deployment quality measured by probabilistic forecasting performance and state turnover. Additional asset classes, prediction tasks, maintenance rules, and prospective live deployment with frozen economic weights and explicit release costs remain future work.

\section{Related Work}
\begin{table}[t]
\caption{\textbf{Positioning of SBS relative to adjacent areas.}}
\label{tab:positioning}
\centering
\footnotesize
\setlength{\tabcolsep}{2.5pt}
\begin{tabular}{p{0.24\columnwidth}p{0.20\columnwidth}p{0.46\columnwidth}}
\toprule
Area & Primary goal & SBS differs because it \\
\midrule
Test-time adaptation & Adapt a model & Governs whether a refit replaces it \\
Continual learning & Learn online & Protects the incumbent updater as state \\
Model selection & Choose a model & Compares recursively evolving states \\
Canary release & Bound rollout & Uses delayed labels entirely off path \\
\bottomrule
\end{tabular}
\end{table}

Table~\ref{tab:positioning} makes the distinction explicit. Forecasting work improves and evaluates the predictor itself through limit-order-book architectures, benchmarks, and microstructural guidance \cite{deeplob,lobcast,lobframe,lit}. Test-time adaptation methods instead update a deployed predictor or its statistics as the input distribution changes \cite{ttt,tent,bnstats}. SBS can govern either kind of candidate, but it does not prescribe its training objective or architecture. It asks a later question: whether the resulting state has earned authority to replace the incumbent that has continued to adapt during fitting.

Retraining policies decide when expected benefit justifies a new fit \cite{regol}. Model-selection methods choose among pretrained candidates using online experiments or actively acquired labels \cite{daiproductionselection,karimiactivemodel}. ChaCha is closer: for online AutoML, it maintains live champion and challenger configurations, promotes using progressive validation bounds, and recursively expands a configuration search \cite{chacha}. A recent finance framework instead optimizes when to switch to a challenger whose evidence improves as new features accumulate \cite{challenger2025}. SBS holds candidate generation and cadence fixed and estimates a delayed-label release effect: a warm-refitted state and the actually maintained incumbent traverse the same bounded off-path trial, and the accepted state initializes the next trial. Its outcome is the complete recursive serving trajectory rather than accuracy at a single checkpoint.

Production-readiness systems separate validation, release, monitoring, and recovery \cite{breckmltest,tfx}; canary analysis limits exposure while measuring a live rollout \cite{beyercanary}. SBS complements those controls by collecting delayed labels off the serving path before any candidate receives traffic. Selective classification likewise withholds action when confidence is insufficient \cite{chowreject,elyanivwiener,geifmanselective}, but its action is an individual prediction. SBS accepts or rejects a model-state transition whose consequences persist across future descendants.

\section{Conclusion}
Retraining and deployment are different decisions. SBS compares a trained challenger with the maintained incumbent on paired forward labels before authorizing replacement. Across two nonoverlapping episodes totaling 48 weeks, it improves on calendar replacement, schedule-matched automatic promotion, and continuous maintenance while avoiding 414 of 528 proposed model changes. Relative to the strong maintenance baseline, the forecast gain accumulates to about 455 total log-loss units per million predictions while deployed-state turnover falls by 78.4\%. Stable effects across seeds, trial budgets, margins, time, assets, and training objectives support the central result. \emph{The contribution is a deployment policy rather than a forecasting architecture}: train often, but require forward evidence before swapping the state that serves.

\sloppy
\bibliographystyle{ACM-Reference-Format}
\bibliography{references}
\end{document}